\documentclass[pra,showpacs,twocolumn,floatfix,aps]{revtex4}
\usepackage{graphicx}
\usepackage{epstopdf}

\usepackage{hyperref}
\usepackage{color}
\usepackage[dvipsnames]{xcolor}
\usepackage{footnote}
\usepackage{subcaption}
\usepackage{physics}
\usepackage{verbatim}
\usepackage{amssymb}

\usepackage{float}
\usepackage{bbm}

\usepackage{caption}

\begin{document}
 \title{
 Enhanced quantum thermometry near a dissipative phase transition in a driven Kerr cavity}
\author{Chayan Purkait}
\email{spscp3458@iacs.res.in,}
\author{Bimalendu Deb}
\email{msbd@iacs.res.in}
\affiliation{
School of Physical Sciences, Indian Association for the
Cultivation of Science, Jadavpur, Kolkata 700032, India
}
\date{\today}%


\begin{abstract}

We investigate quantum thermometry in a driven--dissipative Kerr cavity coupled to a thermal reservoir. The system exhibits a finite-size precursor of a dissipative phase transition, characterized by a pronounced minimum of the Liouvillian gap and a sharp jump in the steady-state physical observables such as average photon number at the critical driving strength. We show that this regime leads to strong enhancement of the quantum Fisher information (QFI) for temperature estimation. Using an effective two-branch description, we show that the enhancement originates from temperature-induced redistribution of the weight factors in photon number distribution between low- and high-photon-number branches, which is described by an effective binary Fisher information. By optimizing the coherent drive, the enhanced response persists over an extended low-temperature, low-thermal-occupation regime and yields a favorable relative temperature-uncertainty bound. These results identify finite-size precursors of dissipative phase transitions in Kerr-cavity platforms as useful resources for tunable nonequilibrium quantum thermometry. We further show that the predicted thermometric enhancement is accessible in a parameter regime compatible with circuit quantum electrodynamics (circuit-QED) platforms.

\end{abstract}

\maketitle

\section{Introduction}


Precise temperature estimation is a central task in quantum science and technology, with applications ranging from the calibration of cryogenic quantum devices to the characterization of nanoscale thermal environments, quantum simulators, and low-temperature sensing platforms \cite{mehboudi2019JPA,binder2018book,potts2019Quantum,scigliuzzo2020PRX}. In the quantum regime, temperature is not associated with a direct observable, but must be inferred indirectly from measurements performed on a quantum probe that interacts with a sample or a thermal bath \cite{binder2018book,mehboudi2019JPA}. Quantum thermometry \cite{doicin2026PRA,rubio2020PRL,tumbiolo2026PRL,montenegro2020PRR,chattopadhyay2025PRA,ullah2025QST,ullah2023PRR,chattopadhyay2026Arxiv,pati2020PRA} combines open quantum-system dynamics with quantum estimation theory, where the ultimate achievable precision is quantified by the quantum Fisher information (QFI) through the quantum Cramér--Rao bound \cite{braunstein1994PRL,paris2009IJQI,binder2018book,mehboudi2019JPA}. The QFI characterizes the distinguishability of probe states associated with infinitesimally different temperatures \cite{braunstein1994PRL,paris2009IJQI}. Enhancing the QFI is therefore a key objective in the design of high-precision quantum thermometers.


A major challenge in quantum thermometry is to identify physical regimes in which small temperature variations produce a large and experimentally accessible change in the probe state  \cite{correa2015PRL,mehboudi2019JPA,binder2018book}. Conventional equilibrium thermometry relies on the temperature dependence of a thermal state, for which the QFI is closely related to energy fluctuations and is typically optimized only within a limited temperature window \cite{correa2015PRL,potts2019Quantum,campbell2018QST}. 
Research efforts to overcome this limitation have given rise to a variety of nonequilibrium and quantum-enhanced strategies, including coherence-assisted probes \cite{ullah2023PRR,frazao2024Ent}, ancilla-assisted protocols \cite{kiilerich2018PRA}, dynamically controlled thermometers \cite{mukherjee2019CP}, non-Markovian probes \cite{zhang2022PRApp,aiache2024PRE}, many-body sensors \cite{hovhannisyan2021PRXQ,aybar2022Quantum}. In particular, systems operating near phase-transition regimes exhibit enhanced susceptibility to small parameter changes in their steady states or dynamical responses. When this susceptibility has a strong temperature dependence and is experimentally accessible, it can provide a route to enhanced thermometric sensitivity \cite{aybar2022Quantum,yu2024PRR,xie2022EPJP}.

Criticality-enhanced quantum thermometry has recently been explored in several complementary settings \cite{aybar2022Quantum,yu2024PRR,xie2022EPJP}. In a dissipative quantum Rabi system, thermometric precision has been shown to be enhanced near the critical point of the normal-to-superradiant phase transition, rather than near the exceptional point of the effective anti-parity-time-symmetric cavity dynamics \cite{xie2022EPJP}. In equilibrium many-body systems, a finite-size scaling theory has demonstrated that the thermometric QFI can be strongly enhanced near continuous quantum phase transitions through the closing of the Hamiltonian energy gap and the associated critical scaling of thermal fluctuations \cite{aybar2022Quantum}. Critical enhancement has also been proposed in phase thermometry of a finite two-dimensional Ising lattice, where a local spin probe acquires temperature-dependent dephasing from critical thermal spin fluctuations of the surrounding lattice \cite{yu2024PRR}. These studies demonstrate that critical regimes can provide useful resources for temperature estimation.

Here, we study quantum thermometry using a driven-dissipative Kerr cavity—a coherently driven nonlinear bosonic mode coupled to a thermal reservoir
\cite{minganti2018PRA,drummond1980JPA,carmichael2015PRX,akhtar2025PRA}. This system has long served as a minimal model for quantum optical bistability and nonlinear cavity dynamics \cite{drummond1980JPA,drummond1981PRA,casteels2017PRA,bartolo2016PRA}. A thermodynamic limit for this nonequilibrium system is defined by introducing a dimensionless effective system-size parameter $\mathcal{N}$ and scaling the Kerr nonlinearity and coherent-drive amplitude as $U_{\mathcal{N}}=$ $U / \mathcal{N}$ and $F_{\mathcal{N}}=\sqrt{\mathcal{N}} F$, respectively. Here, $U$ and $F$ denote reference values of the Kerr nonlinearity and coherent-drive amplitude, respectively. As $\mathcal{N} \rightarrow \infty$, the steady-state photon number becomes extensive, $\left\langle a^{\dagger} a\right\rangle_{\mathrm{ss}} \propto \mathcal{N}$, while the corresponding photon density, $\left\langle a^{\dagger} a\right\rangle_{\mathrm{ss}} / \mathcal{N}$, remains finite. Under this scaling, the mean-field equations exhibit optical bistability
\cite{drummond1980JPA,drummond1981PRA,casteels2017PRA,bartolo2016PRA}. Although the exact quantum steady state remains unique at finite $\mathcal{N}$, it undergoes a first-order dissipative phase transition in the limit $\mathcal{N} \rightarrow \infty$, at which the steady-state photon density changes discontinuously \cite{drummond1980JPA,drummond1981PRA,casteels2017PRA,bartolo2016PRA}.
For a finite effective system size, the sharp transition is replaced by a smooth crossover between low- and high-photon-number steady-state branches at the transition point \cite{casteels2017PRA,vukics2019Quantum,minganti2018PRA}. This crossover is accompanied by a pronounced reduction of the Liouvillian gap, indicating the emergence of slow relaxation dynamics and a finite-size precursor of the underlying dissipative transition \cite{casteels2017PRA,vukics2019Quantum,minganti2018PRA}. 
The coexistence of a sharp steady-state crossover and a small but finite Liouvillian gap makes a driven–dissipative Kerr cavity system an attractive platform for investigating quantum thermometry \cite{correa2015PRL,potts2019Quantum,ullah2023PRR,mok2021CP}. Existing criticality-enhanced thermometry schemes rely on Hamiltonian gap closing in equilibrium many-body systems \cite{aybar2022Quantum}, probe dephasing from critical fluctuations \cite{yu2024PRR}, or second-order dissipative and exceptional-point physics \cite{xie2022EPJP}; none exploit the temperature-driven redistribution of steady-state weight between coexisting photon-number branches near a first-order driven–dissipative transition.

We investigate temperature estimation using the nonequilibrium steady-state QFI and show that the thermometric sensitivity is strongly enhanced near the finite-size precursor of the dissipative phase transition. The optimal thermometric response occurs at a small but finite gap, allowing the system to retain critical-like sensitivity without requiring the divergent relaxation time associated with an exactly closed gap. By optimizing the coherent drive, we further demonstrate that this enhancement persists over an extended low-temperature, low-thermal-occupation regime. Finally, using an effective two-branch description, we show that the enhancement originates from temperature-induced redistribution of weight between low- and high-photon-number branches, rather than from usual thermalization to the bath. We also assess the experimental feasibility of this mechanism in a circuit-QED-compatible parameter regime and find that the close correspondence among the photon-number crossover, Liouvillian-gap minimum, and QFI maximum persists. The predicted enhancement may be probed using microwave-state tomography or classical Fisher information extracted from experimentally accessible output-field measurements.



The paper is organized as follows. In Sec. \ref{model}, we introduce the driven--dissipative Kerr cavity model and the thermal Lindblad master equation. In Sec. \ref{phase}, we discuss the finite-size precursor of the dissipative phase transition, diagnosed through the Liouvillian gap and the steady-state photon number. In Sec. \ref{QFI}, we review the quantum Fisher information formalism for temperature estimation in a nonequilibrium steady state. In Sec. \ref{Rel}, we present our main results: the enhancement of the steady-state QFI near the transition precursor, an effective two-branch description that identifies branch-weight redistribution as the origin of the enhancement, and the optimized low-temperature thermometric performance. In Sec. \ref{Exp}, we discuss experimental feasibility in circuit-QED platforms and possible measurement strategies. Finally, Sec. \ref{Con} summarizes our conclusions.





\begin{figure}[h!]
\includegraphics[width=0.45\textwidth]{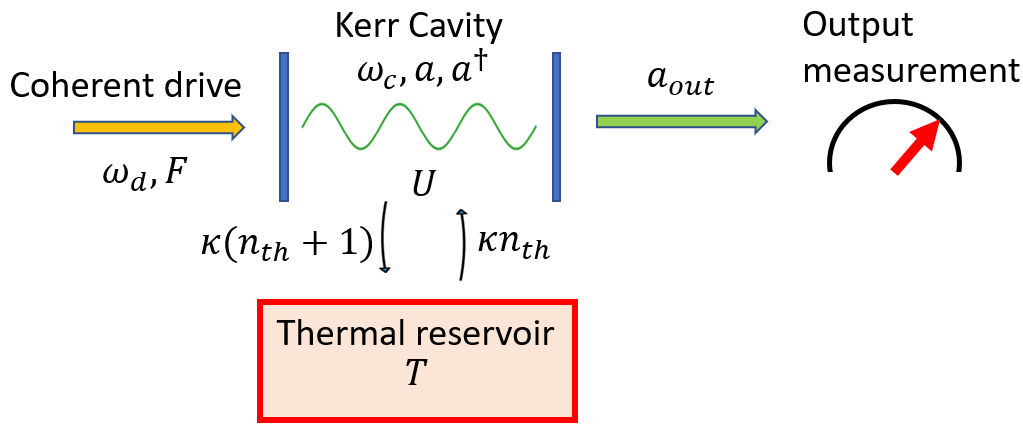}
   \caption{Schematic of the driven-dissipative Kerr cavity used for quantum thermometry. A single nonlinear cavity mode of resonance frequency $\omega_c$, described by the bosonic annihilation and creation operators $a$ and $a^{\dagger}$, is coherently driven at frequency $\omega_d$ with drive amplitude $F$. The cavity possesses a Kerr nonlinearity of strength $U$. It is coupled to a thermal reservoir at temperature $T$, with $\kappa$ denoting the cavity energy-decay rate. The reservoir induces photon-loss and photon-absorption processes with rates $\kappa\left(n_{\text {th }}+1\right)$ and $\kappa n_{\text {th }}$, respectively, where
$
n_{\mathrm{th}}=\left[\exp \left(\frac{\hbar \omega_c}{k_{\mathrm{B}} T}\right)-1\right]^{-1}
$
is the mean thermal occupation of the reservoir at the cavity frequency. The emitted output field, denoted by $a_{\text {out }}$, is monitored to infer the reservoir temperature from the cavity response. Here, $k_{\mathrm{B}}$ and $\hbar$ are the Boltzmann and reduced Planck constants, respectively.} 
   \label{fig:Schematic}
\end{figure}

\section{Model}\label{model}

Our model is schematically shown in Fig.~\ref{fig:Schematic}.
We consider a single-mode driven--dissipative Kerr nonlinear cavity coupled to a thermal reservoir \cite{drummond1980JPA,carmichael2015PRX}. The cavity mode has a frequency \(\omega_c\), with photon annihilation and creation operators \(a\) and \(a^\dagger\). Let \(U\) be the Kerr nonlinearity, and \(F\) be the coherent drive amplitude, then the Hamiltonian in a frame rotating at the drive frequency \(\omega_d\) in the unit of \(\hbar=k_B=1\) is given by
\begin{equation}
H=-\Delta a^\dagger a+\frac{U}{2}a^\dagger a^\dagger a a
+F(a+a^\dagger),
\end{equation}
where \(\Delta=\omega_d-\omega_c\) is the drive--cavity detuning. 

The cavity is coupled to a thermal bath at temperature \(T\). After tracing out the bath degrees-of-freedom, and under the weak-coupling, Markovian, and secular approximations, 
the reduced density matrix of the cavity mode obeys the Lindblad master equation \cite{breuer2002Book}
\begin{equation}
\dot{\rho}=\mathcal{L}[\rho]
=-i[H,\rho]
+\kappa(n_{\rm th}+1)\mathcal{D}[a]\rho
+\kappa n_{\rm th}\mathcal{D}[a^\dagger]\rho ,
\end{equation}
where
$$
\mathcal{D}[O]\rho
=O\rho O^\dagger-\frac{1}{2}\{O^\dagger O,\rho\},
$$
and \(\kappa\) is the cavity decay rate. 

Although the system Hamiltonian is expressed in a frame rotating at the drive frequency, the thermal occupation entering the dissipator is evaluated at $\omega_c$. This follows from the microscopic system--bath theory: absorption and emission rates are determined by the bath correlation functions evaluated at the frequency of the cavity. In particular, the phase acquired by annihilation operator $a$ cancels within the dissipator $\mathcal{D}[a]$. We therefore use 
\begin{equation}
    n_{\rm th}=[\exp(\omega_c/T)-1]^{-1},
\end{equation}
assuming weak Kerr anharmonicity and a bath spectral density that varies negligibly across the relevant cavity linewidth.


The non-equilibrium steady state is defined by
\begin{equation}
\mathcal{L}[\rho_{\rm ss}]=0,\qquad {\rm Tr}[\rho_{\rm ss}]=1.
\end{equation}
For numerical calculations, the infinite-dimensional cavity Hilbert space is truncated to the first \(N\) Fock states, \(\{|0\rangle,\ldots,|N-1\rangle\}\). The cutoff \(N\) is chosen sufficiently large to ensure convergence of the steady-state photon number, Liouvillian gap, and quantum Fisher information. In our numerical work, we set $N=60$, which is found to be sufficiently large to ensure convergence.

The Liouvillian \(\mathcal{L}\) is a linear, generally non-Hermitian superoperator acting on the space of density matrices. Its eigenvalues \(\lambda_k\) determine the relaxation dynamics. For stable Lindblad evolution, \({\rm Re}(\lambda_k)\leq 0\), and the zero eigenvalue corresponds to the steady state. 






\begin{figure}[h!]
\includegraphics[width=0.45\textwidth]{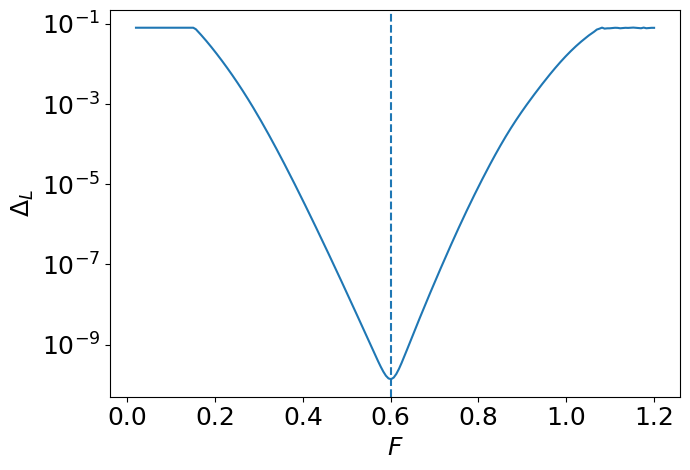}
   \caption{The variation of Liouvillian gap $\Delta_L$ as a function of drive amplitude $F$. The parameters are $\omega_c = 1.0$, $\kappa = 0.08$, $U = 0.03$, $\Delta = 0.6$, $T = 0.15$.  } 
   \label{fig:Liouvillian Gap Vs Drive Amplitude}
\end{figure}

\section{Dissipative Phase Transitions and Their Finite-Size Precursors}\label{phase}

\begin{figure}[h!]
\includegraphics[width=0.45\textwidth]{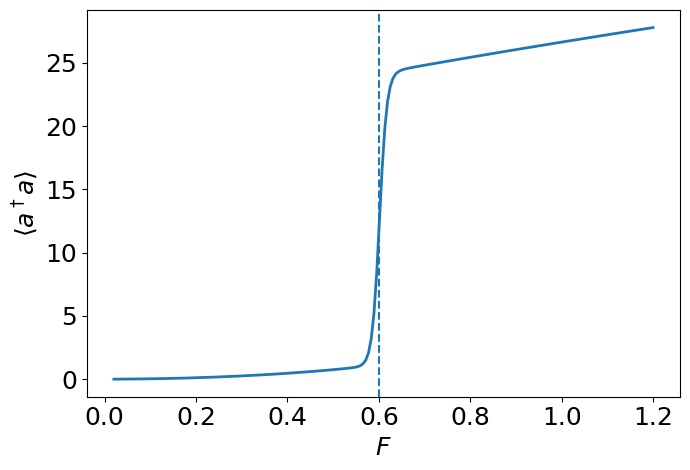}
   \caption{The variation of average photon number $\langle a^{\dagger}a\rangle$ in the steady state as a function $F$. The other  parameters are same as in Fig.~\ref{fig:Liouvillian Gap Vs Drive Amplitude}. } 
   \label{fig:Photon number Vs Drive Amplitude}
\end{figure}

A dissipative phase transition (DPT)
\cite{minganti2018PRA,sieberer2016RPP,carmichael2015PRX} is the non-equilibrium analogue of an equilibrium phase transition in an open quantum system governed by a Liouvillian superoperator
\(\mathcal{L}(\lambda)\). In the thermodynamic limit, the
nonequilibrium steady state \(\rho_{\rm ss}(\lambda)\), defined by \(\mathcal{L}(\lambda)\rho_{\rm ss}=0\), can become nonanalytic as a function of a control parameter \(\lambda\), such as the drive strength, detuning, or interaction strength. This nonanalyticity appears through abrupt or singular changes in steady-state observables, analogous to
first- and second-order equilibrium phase transitions. From the spectral viewpoint, a steady-state DPT is typically associated with the closing of the Liouvillian gap,
\begin{equation}
   \Delta_L=-{\rm Re}(\lambda_1)\rightarrow 0, 
\end{equation}
where \(\lambda_1\) is the nonzero eigenvalue whose real part is closest
to zero. The closing of this gap implies a divergent relaxation time,
\(\tau\sim 1/\Delta_L\) \cite{minganti2018PRA}, and gives rise to
critical slowing down and long-lived Liouvillian modes. 


For finite systems, the steady state of a Markovian open quantum system is generally unique and analytic in the control parameter, so a true nonanalytic phase transition is absent \cite{minganti2018PRA,macieszczak2016PRL,carmichael2015PRX}. Instead, the transition is replaced by a sharp but smooth crossover. In this sense, a finite system does not exhibit a genuine dissipative phase transition, but rather a finite-size precursor of the transition that emerges in the thermodynamic limit. A useful operational signature of such a precursor is provided by the Liouvillian spectrum. Near this pseudo-critical region, the Liouvillian gap becomes small but remains finite, producing long relaxation times and a finite-size precursor of critical slowing down \cite{minganti2018PRA,macieszczak2016PRL}. This small-gap region signals a metastable Liouvillian structure, in which one or a few slow modes are well separated from the rapidly decaying modes. The associated metastable states may be interpreted as finite-size remnants of the coexisting phases that emerge in the thermodynamic limit. They can give rise to switching behavior, enhanced fluctuations, and bimodal distributions of suitable order parameters, thereby providing finite-size precursors of the DPT.

\subsection{Dissipative Phase Transition in a Kerr Nonlinear Cavity}



We now illustrate the finite-size precursor of a DPT in the driven--dissipative Kerr cavity introduced in Sec.~\ref{model}.
The spectral signature of the crossover associated with a DPT is shown in Fig.~\ref{fig:Liouvillian Gap Vs Drive Amplitude}, where we plot the Liouvillian gap \(\Delta_L=-{\rm Re}(\lambda_1)\) as a function of the drive amplitude \(F\). Note that, for our numerical works, all frequency parameters are scaled by $\omega_c$. As \(F\) approaches the pseudo-critical region, the gap decreases by several orders of magnitude, indicating the emergence of a slow relaxation mode. The corresponding relaxation time, \(\tau\sim 1/\Delta_L\), therefore becomes large, which is the finite-size manifestation of critical slowing down. 
This small-gap region indicates that the long-time relaxation dynamics is dominated by a slowly decaying Liouvillian mode. 


In Fig.~\ref{fig:Photon number Vs Drive Amplitude}, we plot steady-state photon number \(\langle a^\dagger a\rangle_{\rm ss}\) as a function of \(F\). The photon number exhibits a sharp but continuous crossover from a low-occupation branch at weak driving to a high-occupation branch at stronger driving. This behavior is the finite-size remnant of nonlinear optical bistability, which becomes a discontinuous jump in the thermodynamic scaling limit. The crossover region in Fig.~\ref{fig:Photon number Vs Drive Amplitude} occurs close to the minimum of the Liouvillian gap shown in Fig.~\ref{fig:Liouvillian Gap Vs Drive Amplitude}, indicating that the rapid restructuring of the steady state is correlated with the emergence of slow Liouvillian relaxation dynamics. Thus, 
the driven Kerr cavity displays the key finite-size signatures of a dissipative phase transition: a pronounced minimum of the Liouvillian gap, a sharp crossover in the steady-state photon number, and a rapid restructuring of the nonequilibrium steady state between low- and high-photon-number regimes. 
As discussed in the following section, this enhanced steady-state susceptibility provides the physical basis for the amplification of temperature sensitivity.

\section{Quantum Fisher Information for Temperature Estimation}\label{QFI}

Quantum thermometry aims to estimate the temperature \(T\) of an
environment by measuring a quantum probe whose state depends on \(T\)
\cite{mehboudi2019JPA}. The ultimate precision of any unbiased estimator
is bounded by the quantum Cramér--Rao inequality
\cite{braunstein1994PRL,paris2009IJQI},
\begin{equation}
\delta T \geq \frac{1}{\sqrt{\nu F_Q(T)}},
\end{equation}
where \(\delta T\) is the standard deviation of the estimator, \(\nu\)
is the number of independent measurements, and \(F_Q(T)\) is the quantum
Fisher information (QFI).

The QFI quantifies the distinguishability of neighboring probe states
\(\rho(T)\) and \(\rho(T+dT)\). It is defined as
\begin{equation}
F_Q(T)={\rm Tr}\left[\rho(T)(L_T^{\rm SLD})^2\right],
\end{equation}
where \(L_T^{\rm SLD}\) is the symmetric logarithmic derivative,
determined by
\begin{equation}
\partial_T \rho(T)=\frac{1}{2}
\left(L_T^{\rm SLD}\rho(T)+\rho(T)L_T^{\rm SLD}\right).
\end{equation}
For a mixed state with spectral decomposition
\begin{equation}
\rho(T)=\sum_n p_n(T)|n(T)\rangle\langle n(T)|,
\end{equation}
the QFI can be written as
\begin{equation}
F_Q(T)=
\sum_n \frac{[\partial_T p_n]^2}{p_n}
+
2\sum_{n\neq m}
\frac{(p_n-p_m)^2}{p_n+p_m}
|\langle n|\partial_T m\rangle|^2 ,
\end{equation}
where terms with \(p_n+p_m=0\) are omitted.
The first term is the classical, or population, contribution and arises
from the temperature dependence of the eigenvalues \(p_n\). The second
term is the quantum contribution and originates from the temperature
dependence of the eigenvectors.


For an equilibrium Gibbs state with a temperature-independent Hamiltonian, thermometric sensitivity is directly related to energy fluctuations. By contrast, for a nonequilibrium steady state, the QFI depends on the full temperature dependence of $\rho_{\rm ss}(T)$. Near a finite-size precursor of a dissipative phase transition, this dependence can be strongly amplified by slow Liouvillian dynamics and rapid steady-state restructuring.

\begin{figure}[h!]
\includegraphics[width=0.45\textwidth]{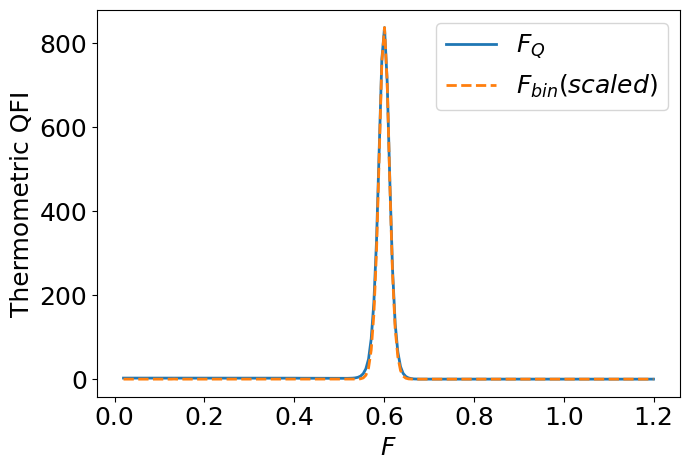}
   \caption{The variation of quantum Fisher information $F_Q$ (blue color solid line) and $F_{bin}$ (orange color dashed line) as a function of drive amplitude $F$. The other parameters are same as in Fig.~\ref{fig:Liouvillian Gap Vs Drive Amplitude}.} 
   \label{fig:QFI Vs F with two state prediction}
\end{figure}

\section{Results and Discussions}\label{Rel}

\subsection{Enhanced quantum Fisher information}

We now analyze the steady-state QFI for temperature estimation in our system. In Fig.~\ref{fig:QFI Vs F with two state prediction}, we plot \(F_Q(T)\) as a function of the drive amplitude \(F\). A striking feature is the emergence of a pronounced peak in the QFI in the vicinity of the crossover region identified in Sec.~\ref{phase}. This region lies close to the minimum of the Liouvillian gap (Fig.~\ref{fig:Liouvillian Gap Vs Drive Amplitude}) and to the sharp variation of the steady-state photon number (Fig.~\ref{fig:Photon number Vs Drive Amplitude}), indicating that the enhancement of thermometric sensitivity is closely associated with the finite-size precursor of the DPT.


The full QFI is evaluated from the steady-state density matrix \(\rho_{\rm ss}(T)\). Near the finite-size transition precursor, the dominant temperature dependence is found to arise mainly from changes in
the eigenvalue distribution of \(\rho_{\rm ss}\), rather than from changes in its eigenvectors. Consequently, the QFI is largely governed by the
population contribution,
\begin{equation}
F_Q(T)\simeq \sum_{p_n>0} \frac{[\partial_T p_n]^2}{p_n}.
\end{equation}
This indicates that the thermometric enhancement is primarily driven by temperature-induced redistribution of population within the steady state. Contributions from changes in the eigenvectors, coherences are comparatively subleading. 
This observation provides the basis for the reduced two-branch description introduced below.


The optimal thermometric response is found to be 
with the finite-size pseudo-critical region where the steady state is
highly sensitive to temperature variations while the system remains
dynamically accessible. This is a key distinction from an ideal
critical-point scenario: a closing Liouvillian gap would lead to
divergent relaxation times, whereas here the finite gap allows one to
exploit critical-like sensitivity without requiring infinitely long
relaxation times.

\begin{figure*}[t]
\includegraphics[width=1\textwidth]{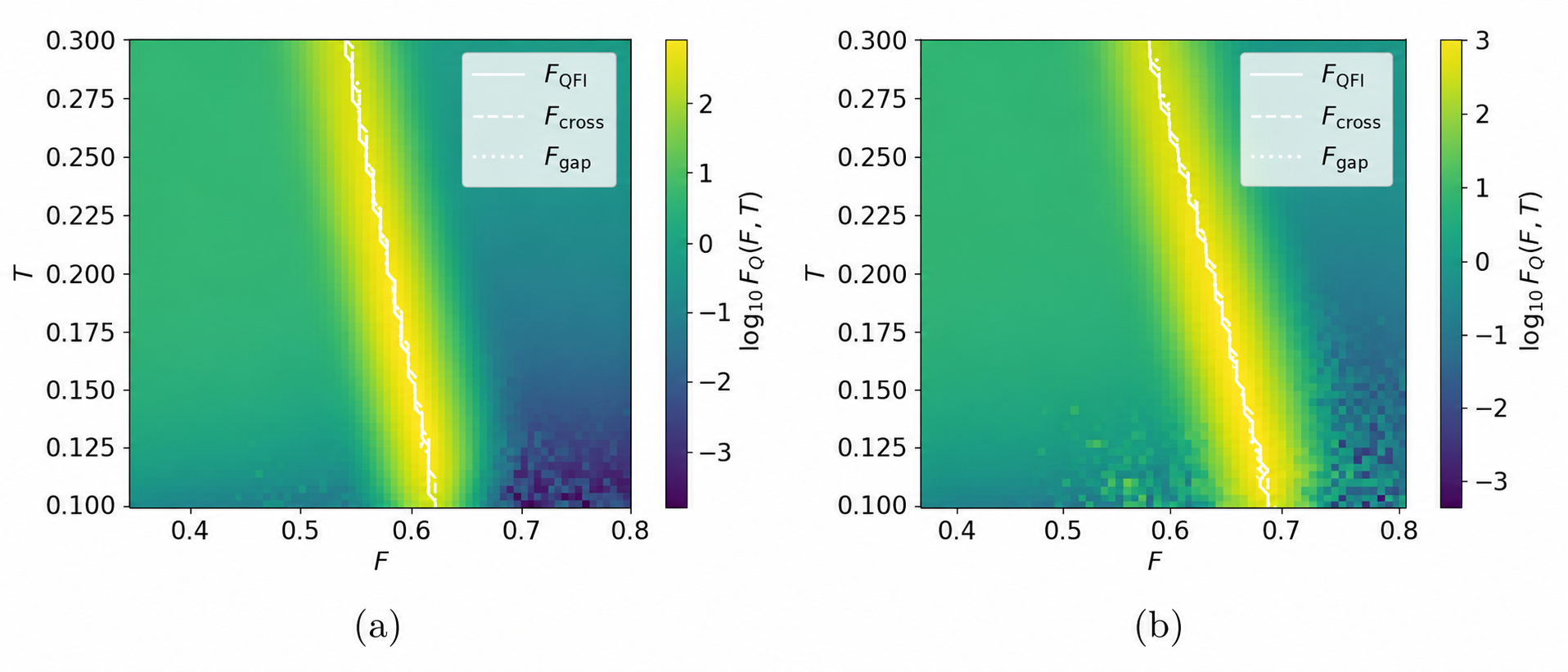}
   \caption{ Quantum Fisher information $F_Q(F, T)$ of the nonequilibrium steady state $\rho_{\mathrm{ss}}(F, T)$ in the drive-temperature plane. Panels (a) and (b) correspond to $\Delta = 0.6$ and $\Delta = 0.65$. The other parameters are the same as in Fig.~\ref{fig:Liouvillian Gap Vs Drive Amplitude}. The solid white curve denotes the location of the maximum QFI at each temperature,
$ F_{\mathrm{QFI}}(T)=\underset{F}{\arg \max } F_Q(F, T)$. The dashed white curve denotes the photon-number crossover line $
F_{\text {cross }}(T)=\underset{F}{\arg \max }\left|\partial_F\left\langle a^{\dagger} a\right\rangle_{\mathrm{ss}}\right|
$,
while the dotted white curve indicates the minimum-Liouvillian-gap line
$
F_{\text {gap }}(T)=\underset{F}{\arg \min } \Delta_L(F, T)
$. The enhanced-QFI ridge closely follows the photon-number crossover and minimum-gap lines, showing that the strongest thermometric sensitivity is concentrated near the finite-size precursor of the critical region. } 
   \label{fig:QFI surface plot}
\end{figure*}



To verify that the enhancement of thermometric sensitivity is not restricted to a single parameter point, we compute the steady-state QFI over the two-dimensional parameter space spanned by the drive amplitude \(F\) and the bath temperature \(T\). For each value of \(T\), we identify the drive amplitude
\[
F_{\rm QFI}(T)=\arg\max_F F_Q(F,T),
\]
at which the QFI is maximal, where, $\arg \max _F$ denotes the set of values of $F$ that maximize $F_Q(F, T)$ at fixed $T$. We then compare this optimal-sensitivity points
with the photon-number crossover point,
\[
F_{\rm cross}(T)=\arg\max_F 
\left|\partial_F \langle a^\dagger a\rangle_{\rm ss}\right|,
\]
and with the minimum-gap point,
\[
F_{\rm gap}(T)=\arg\min_F \Delta_L(F,T).
\]


Fig.~\ref{fig:QFI surface plot} shows that the QFI forms a pronounced ridge in the $(F,T)$ plane. This ridge closely follows the photon-number crossover line and remains near the minimum-gap line. Thus, the optimal thermometric response occurs in the finite-size precursor region of the dissipative transition. 
The QFI maximum has closer alignment with the photon-number crossover than with the precise minimum of the Liouvillian gap, suggesting that the enhancement is more directly connected to the rapid restructuring of the steady state rather than to the gap minimum alone. This connection is analyzed quantitatively in the effective two-branch description introduced below.


Increasing the detuning from $\Delta=0.6$ to $\Delta=0.65$ modifies the nonlinear cavity response and shifts the finite-size crossover region in the $(F,T)$ plane. Consequently, the maximum-QFI, photon-number crossover, and minimum-gap lines are displaced in drive amplitude, and their quantitative separation is also modified. Nevertheless, their qualitative relationship remains unchanged: the QFI ridge continues to closely follow the photon-number crossover and remains in the vicinity of the minimum Liouvillian gap. These results therefore serve as a check for the robustness of our proposed quantum thermometry demonstrating that the thermometric enhancement is not required to be fine-tuned to a single detuning only. 

\subsection{Effective two-branch model}

The physical origin of the QFI enhancement can be traced to the strong temperature dependence of the nonequilibrium steady state near the finite-size transition precursor. In this regime, the steady state undergoes a rapid restructuring between low- and high-photon-number branches and its leading temperature dependence can be captured by a temperature-dependent branch weight. It is therefore useful to approximate the steady state as a temperature-dependent mixture of effective two-branch contributions. It provides an operational reduced description based on the observed low- and high-photon-number branches of the steady state. In this effective two-branch description, the nonequilibrium steady state is written as
\begin{equation}
\rho_{\rm ss}(T) \simeq p(T)\,\rho_+ + [1-p(T)]\,\rho_- ,
\end{equation}
where \(\rho_+\) and \(\rho_-\) denote representative density matrices
associated with the high- and low-photon-number branches, respectively. The parameter \(p(T)\) is the temperature-dependent weight of the
high-photon-number branch.  Within this approximation, the leading contribution to
\(\partial_T \rho_{\rm ss}\) is attributed to the temperature dependence of
\(p(T)\), so that
\[
\partial_T \rho_{\rm ss}(T) \simeq [\partial_T p(T)](\rho_+ - \rho_-).
\]
A small variation of the bath temperature can then produce a pronounced redistribution of weight between these branches, leading to a large change in the steady-state density matrix and, consequently, to an enhanced QFI.

\begin{figure*}[t]
\includegraphics[width=1\textwidth]{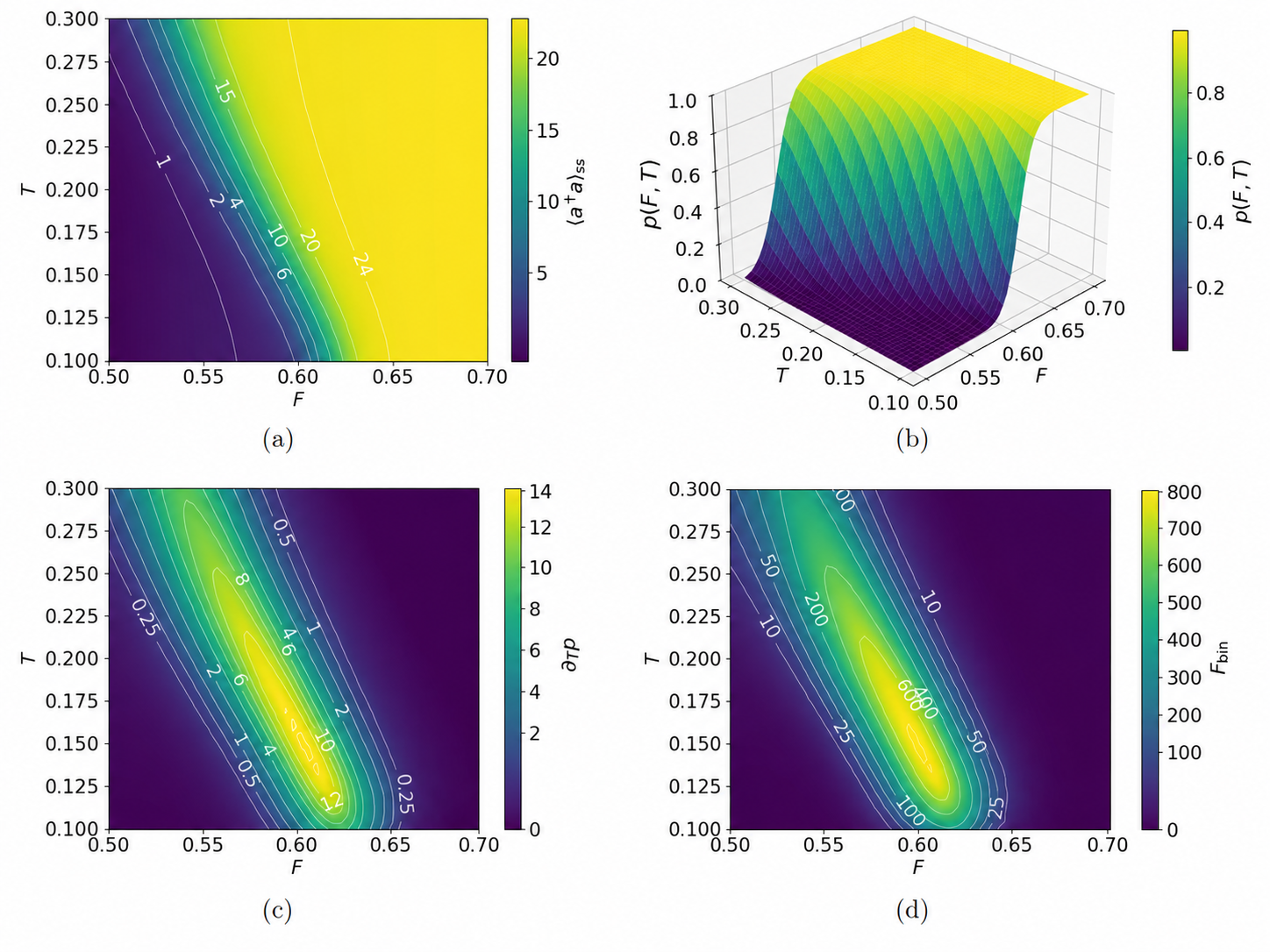}
   \caption{ Branch-occupation diagnostics in the drive--temperature plane.
    (a) Steady-state photon number \(\langle a^\dagger a\rangle_{\rm ss}(F,T)\)
    as a function of drive amplitude \(F\) and temperature \(T\). 
    (b) Three-dimensional reconstruction of the effective high-photon-number branch occupation $p(F,T)$, obtained from Eq.~\ref{brach occupation}.
    (c) Temperature derivative $\partial_T p(F,T)$.
    (d) Binary Fisher-information contribution
    $F_{\rm bin}(F,T)$.  The other parameters are same as in Fig.~\ref{fig:Liouvillian Gap Vs Drive Amplitude}. } \label{fig:branch_occupation_diagnostics}
\end{figure*}

To extract the effective branch weight from the full steady state, we use the steady-state average photon number $\langle a^\dagger a\rangle_{\rm ss}$ as an order parameter. Since the low- and high-photon-number branches are defined by the driven Kerr response as a function of the drive strength, the branch reconstruction is performed at fixed temperature. For each value of $T$, the photon-number curve $\langle a^\dagger a\rangle_{\rm ss}(F,T)$ is fitted away from the crossover region, where the steady state is predominantly localized on either the low- or high-photon-number branch (see Fig.~\ref{fig:Photon number Vs Drive Amplitude}). This procedure yields the effective low- and high-photon-number branch values, denoted by $n_-(F,T)$ and $n_+(F,T)$, respectively. The steady-state photon number is then approximated as
\begin{equation}
\langle a^\dagger a\rangle_{\rm ss}(F,T)
\simeq p(F,T)n_+(F,T)+[1-p(F,T)]n_-(F,T).
\end{equation}
This relation allows us to reconstruct the effective occupation of the high-photon-number branch as
\begin{equation}\label{brach occupation}
p(F,T)=
\frac{\langle a^\dagger a\rangle_{\rm ss}(F,T)-n_-(F,T)}
{n_+(F,T)-n_-(F,T)}.
\end{equation}
The temperature derivative $\partial_T p(F,T)$ is evaluated numerically using a central finite-difference scheme.

Within this effective binary description, the dominant temperature dependence of the steady state is captured by the redistribution of weight between two effective photon-number branches. The QFI is therefore mainly governed by the population contribution and can be approximated by the Fisher information of a binary distribution,
\begin{equation}\label{TBP}
F_{\rm bin}(F,T)=
\frac{[\partial_T p(F,T)]^2}
{p(F,T)[1-p(F,T)]}.
\end{equation}
The effective two-branch description shows that the QFI enhancement is controlled by two key ingredients. First, the branch weight \(p(T)\) must be highly sensitive to temperature, which occurs near the steady-state crossover region. Second, both low- and high-photon-number branches must contribute appreciably to the steady state, corresponding to the intermediate regime \(0<p<1\). Near the finite-size transition precursor, small variations in the bath temperature produce rapid changes in \(p(T)\) over a narrow parameter range. At the same time, the simultaneous contribution of both branches prevents the binary Fisher information from being suppressed by vanishing probabilities. Together, these two features lead to a pronounced peak in the thermometric sensitivity.

These features are explicitly demonstrated in Fig.~\ref{fig:branch_occupation_diagnostics}. Fig.~\ref{fig:branch_occupation_diagnostics}(a) shows the steady-state photon number $\langle a^\dagger a\rangle_{\rm ss}$ in the drive--temperature plane. The sharp crossover from the low-photon-number regime to the high-photon-number regime shifts systematically toward lower drive amplitudes as the temperature is increased, indicating that the steady-state restructuring is strongly temperature dependent. Using this photon-number response, we reconstruct the effective high-photon-number branch occupation $p(F,T)$ from Eq.~\ref{brach occupation}. The resulting three-dimensional surface is shown in Fig.~\ref{fig:branch_occupation_diagnostics}(b). Away from the crossover region, $p(F,T)$ remains close to either $p\simeq 0$, where the low-photon-number branch dominates, or $p\simeq 1$, where the high-photon-number branch dominates. In the crossover region, however, $p(F,T)$ changes rapidly between these limiting values, 
demonstrating that small temperature variations can induce a pronounced redistribution of the steady-state weight between the low- and high-photon-number branches. The corresponding temperature derivative, $\partial_T p(F,T)$, shown in Fig.~\ref{fig:branch_occupation_diagnostics}(c), exhibits a pronounced ridge in the drive--temperature plane. This ridge identifies the parameter region in which the effective branch occupation is most sensitive to temperature variations. The same region also gives rise to an enhanced binary Fisher-information contribution $F_{\rm bin}(F,T)$, as shown in Fig.~\ref{fig:branch_occupation_diagnostics}(d). The close correspondence among the photon-number crossover, the rapid variation of the reconstructed branch occupation, the ridge in $\partial_Tp(F,T)$, and the enhancement of $F_{\rm bin}(F,T)$ confirms that the enhanced thermometric response originates predominantly from temperature-sensitive redistribution of the steady state between the two effective photon-number branches near the finite-size transition precursor.

Representative one-dimensional temperature cuts of $p(F,T)$, $\partial_Tp(F,T)$, and the corresponding binary Fisher information for selected drive amplitudes are presented in Appendix Fig.~\ref{fig:branch-occupation-diagnostics-line-plots}. These cuts provide a more direct view of how the temperature-sensitive branch redistribution and the associated Fisher-information peak shift as the coherent drive is varied.

In Fig.~\ref{fig:QFI Vs F with two state prediction}, we compare the full steady-state QFI, computed from the full numerical steady-state density matrix, with the prediction of the effective two-branch model. The two-branch prediction is rescaled by a constant factor to match the peak amplitude of the full QFI, allowing a direct comparison of the peak position and shape without emphasizing the overall amplitude mismatch. The effective model accurately reproduces both the position and the sharpness of the QFI peak. This agreement is notable because the reconstruction of $p(F,T)$ is based only on the steady-state photon number and therefore captures the dominant branch redistribution rather than the full density-matrix structure. The unscaled comparison is shown in Appendix Fig.~\ref{fig:Two State Mechanism Comparison}, where the peak position and overall structure are already captured despite a difference in amplitude. Taken together, these results demonstrate that the thermometric enhancement is predominantly governed by the temperature-sensitive redistribution of weight between the low- and high-photon-number branches, while residual effects beyond the idealized two-branch description mainly affect the overall magnitude. The nonequilibrium character of the steady state near the QFI peak is
further examined through a trace-distance analysis presented in
Fig.~\ref{fig:Steady State} of
Appendix~\ref{Trace dist}.

\subsection{Low-temperature thermometric performance}

We now examine the thermometric performance in the regime $T/\omega_c \ll 1$. In this range, the mean thermal occupation of the cavity bath,
$$
n_{\rm th}=\left(e^{\omega_c/T}-1\right)^{-1},
$$
remains much smaller than unity, so that the cavity operates in a low-thermal-occupation quantum regime. For the interval $T/\omega_c=0.1\text{--}0.3$ considered below, $n_{\rm th}$ varies from approximately $4.5\times10^{-5}$ to $3.7\times10^{-2}$.

\begin{figure}[h!]
\includegraphics[width=0.45\textwidth]{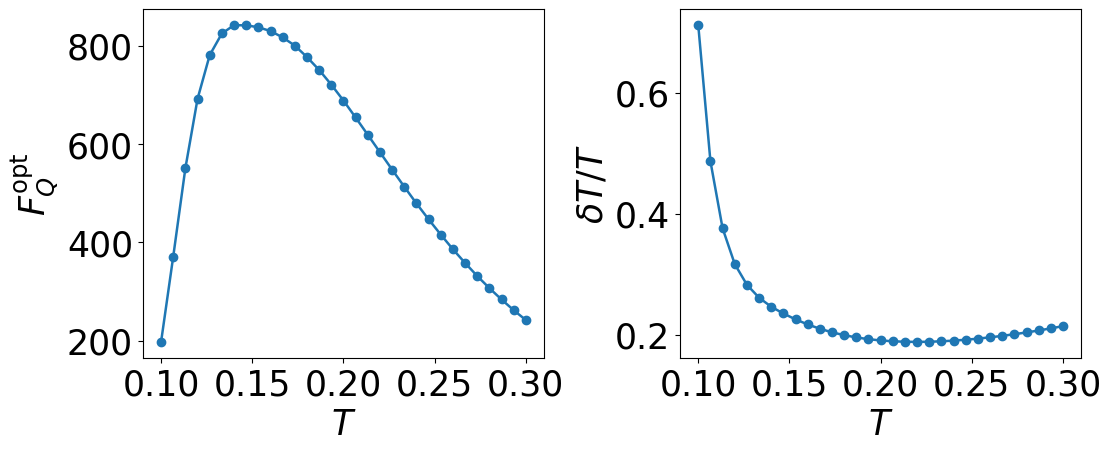}
   \caption{(a) Optimized steady-state quantum Fisher information $F_Q^{\mathrm{opt}}(T)$ as a function of the dimensionless bath temperature $T$. (b) Corresponding single-shot relative temperature-uncertainty bound, $\delta T/T\geq 1/[T\sqrt{F_Q^{\mathrm{opt}}(T)}]$, evaluated for $\nu=1$. The other parameters are the same with Fig.~\ref{fig:Liouvillian Gap Vs Drive Amplitude}.} 
   \label{fig:relative uncertianty}
\end{figure}

To quantify the thermometric performance in the low-temperature regime, we optimize the steady-state quantum Fisher information over the coherent drive amplitude. For each bath temperature \(T\), we define
$$
F_Q^{\mathrm{opt}}(T)
=
\max_F F_Q(F,T),
$$
where the maximization is performed over the drive amplitudes spanning the finite-size crossover region. This optimized quantity represents the largest amount of temperature information that can be encoded in the nonequilibrium steady state when the drive is adjusted to the most sensitive operating point.

The corresponding ultimate precision is determined by the quantum Cramér--Rao bound,
\begin{equation}
\delta T
\geq
\frac{1}{\sqrt{\nu F_Q^{\mathrm{opt}}(T)}},
\end{equation}
where \(\nu\) is the number of independent measurements. Since the same absolute uncertainty has a different significance at different temperatures, it is useful to consider the relative uncertainty,
\begin{equation}
\frac{\delta T}{T}
\geq
\frac{1}{T\sqrt{\nu F_Q^{\mathrm{opt}}(T)}}.
\end{equation}
In the following, we present the single-shot bound corresponding to \(\nu=1\).

Fig.~\ref{fig:relative uncertianty}(a) shows the optimized QFI as a function of the dimensionless temperature \(T/\omega_c\). Starting from the lowest temperature considered, the optimized QFI increases rapidly and reaches a broad maximum around \(T/\omega_c \simeq 0.14\)--\(0.15\), where \(F_Q^{\mathrm{opt}}\) is of the order of \(8\times10^2\). At higher temperatures, the optimized QFI decreases gradually, but remains appreciable throughout the interval \(T/\omega_c=0.1\)--\(0.3\). This behavior shows that the enhancement is not confined to a single temperature point, but persists over an extended low-temperature window when the drive is tuned to the temperature-dependent crossover region.

The corresponding single-shot relative uncertainty is shown in Fig.~\ref{fig:relative uncertianty}(b). Although the optimized QFI is maximal near \(T/\omega_c\simeq0.15\), the minimum relative uncertainty occurs at a somewhat higher temperature, around \(T/\omega_c\simeq0.21\)--\(0.23\). This shift arises from the additional factor of \(T\) in the denominator of the relative-error bound. The minimum value is approximately
\begin{equation}
\left(\frac{\delta T}{T}\right)_{\min}
\simeq 0.19,
\end{equation}
indicating a single-shot relative uncertainty of about \(19\%\). For repeated independent measurements, this bound decreases as \(1/\sqrt{\nu}\); for example, \(\nu=100\) measurements would reduce the corresponding relative uncertainty to approximately \(1.9\%\).


These results demonstrate that the finite-size dissipative-transition precursor produces a pronounced enhancement of the optimized QFI and yields a favorable relative temperature-uncertainty bound within the low-temperature regime considered. The optimal operating temperature is determined by the interplay between the temperature dependence of the optimized QFI and the explicit factor of $T$ entering the relative-error bound. These findings indicate that a driven Kerr cavity can function as a tunable low-temperature thermometric probe, with its ultimate sensitivity optimized by adjusting the coherent drive to track the temperature-dependent crossover region.

\begin{figure*}[t]
\includegraphics[width=0.8\textwidth]{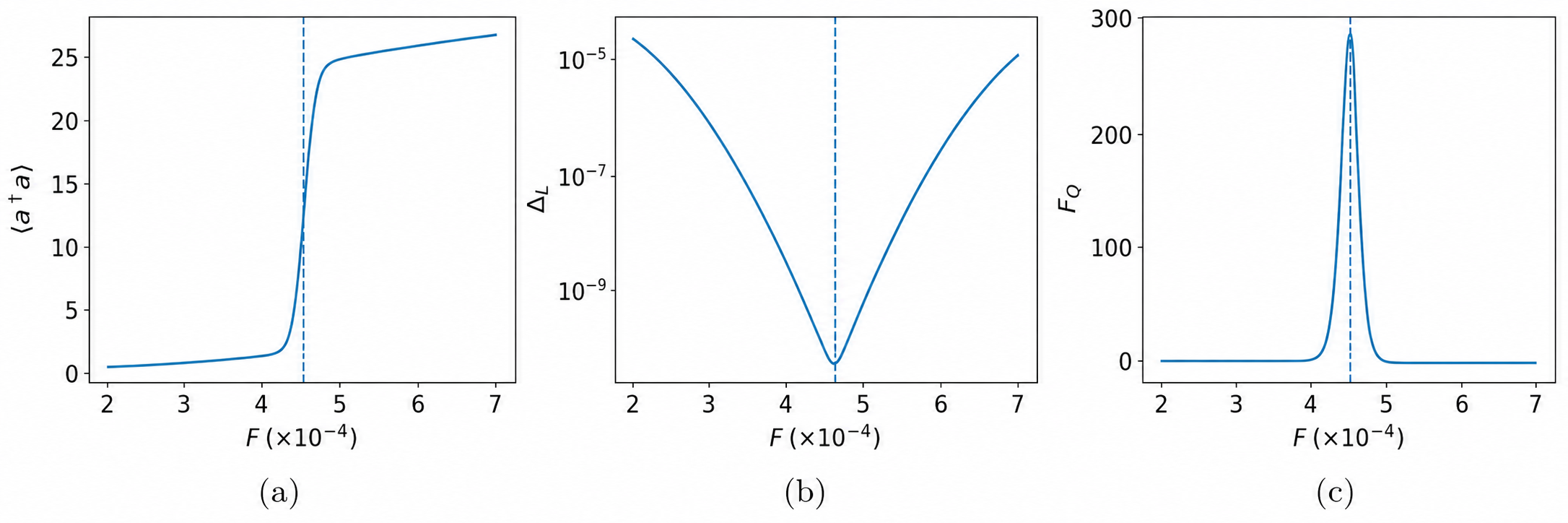}
   \caption{ Thermometric enhancement in a circuit-QED-compatible parameter regime: 
(a) photon number $\langle a^\dagger a\rangle_{\mathrm{ss}}$, 
(b) Liouvillian gap $\Delta_{\mathrm L}$, and 
(c) temperature quantum Fisher information $F_Q$, plotted as functions of drive amplitude $F$. 
The parameters are $\omega_c=1$, $\kappa/\omega_c=10^{-4}$, $U/\kappa=-0.2$, $\Delta/\kappa=-4$, $T/\omega_c=0.2$, and $N=70$. 
The dashed vertical line indicates the minimum-gap point near $F/\kappa\simeq4.55$, which closely coincides with the sharp photon-number crossover and the QFI maximum. } 
   \label{fig:exp data}
\end{figure*}

\section{Experimental feasibility}\label{Exp}

The driven--dissipative Kerr cavity considered in this work can be realized in several state-of-the-art experimental platforms, including superconducting circuit QED \cite{blais2021RMP,kirchmair2013Nature,bourassa2012PRA}, optical Kerr microresonators \cite{Kippenberg2018Science}, exciton-polariton microcavities \cite{carusotto2013RMP}, and other nonlinear optical cavities \cite{lugiato1987PRL}. In these systems, a single bosonic mode with an effective Kerr nonlinearity is coherently driven and coupled to a dissipative environment. 
Among these platforms, superconducting circuit QED is particularly well suited for the present thermometric proposal because it offers strong and tunable nonlinearities, controlled microwave driving, engineered dissipation, direct access to thermal microwave occupations, and high-efficiency measurement of the emitted field \cite{blais2021RMP,kirchmair2013Nature,bourassa2012PRA}.

Driven--dissipative Kerr resonators based on superconducting circuits have been used to observe dissipative phase-transition phenomena, including phase coexistence, hysteresis, symmetry breaking, and critical slowing down \cite{beaulieu2025NC}. Related critical behavior has recently been exploited in a frequency-estimation protocol using a parametric superconducting resonator \cite{beaulieu2025PRXQ}. Temperature-dependent microwave readout in superconducting circuits is experimentally feasible, as demonstrated by dispersive thermometry using a Josephson junction coupled to a resonator \cite{saira2016PRApp}. In the present proposal, this temperature dependence would be inferred from the steady-state response of the driven Kerr resonator. 
The thermometric enhancement may be probed through microwave-state tomography or through classical Fisher information extracted from output-field quadrature distributions, photon-number statistics, branch occupations, and switching dynamics \cite{mallet2011PRL,kirchmair2013Nature}.

Experimentally, one would tune the coherent drive amplitude across the bistable crossover region while maintaining a controlled thermal microwave occupation of the environment. For each bath temperature, the steady-state output field of the resonator could be measured using homodyne or heterodyne detection  \cite{mallet2011PRL,kirchmair2013Nature,blais2021RMP}. The resulting quadrature distributions provide an experimentally accessible measurement record whose temperature dependence can be used to construct a classical Fisher information \cite{braunstein1994PRL,paris2009IJQI,gammelmark2014PRL}. Photon-number statistics, switching trajectories, or inferred branch occupations may also be used for the same purpose \cite{kirchmair2013Nature,beaulieu2025NC}. While the QFI gives the ultimate precision bound and is generally not measured directly, the corresponding classical Fisher information is bounded above by the QFI  \cite{braunstein1994PRL}. Therefore, a pronounced increase of the classical Fisher information near the crossover would provide an experimentally accessible signature of the thermometric enhancement discussed in this work.

To verify that the central mechanism also persists in a circuit-QED-compatible regime, we repeated the calculation using
$\kappa/\omega_c=10^{-4}$, $U/\kappa=-0.2$, and
$\Delta/\kappa=-4$. The results are shown in Fig.~\ref{fig:exp data}. The negative values of $U$ and $\Delta$
correspond, under the convention $\Delta=\omega_d-\omega_c$, to the
usual softening Kerr nonlinearity of a Josephson resonator driven below
resonance. Although this parameter set differs quantitatively from the
benchmark values employed in the main analysis, both regimes satisfy
$|\Delta/U|=20$ and support a well-resolved many-photon crossover.
Scanning the drive over $F/\kappa=2\text{--}7$, we find a sharp
photon-number crossover near $F/\kappa\simeq4.55$, accompanied by a pronounced minimum of the Liouvillian gap and a peak in the steady-state temperature QFI. Taking a representative superconducting-resonator frequency $\omega_c/2\pi=7.15~\mathrm{GHz}$, within the experimentally
demonstrated range of Ref.~\cite{chen2023NC}, this dimensionless
parameter set corresponds to
\[
\frac{\kappa}{2\pi}\simeq0.715~\mathrm{MHz},~
\frac{U}{2\pi}\simeq-0.143~\mathrm{MHz},~
\frac{\Delta}{2\pi}\simeq-2.86~\mathrm{MHz}.
\]
The drive interval $F/\kappa=2\text{--}7$ then corresponds to
$F/2\pi\simeq1.43\text{--}5.01~\mathrm{MHz}$, with the crossover
occurring near $F_c/2\pi\simeq3.25~\mathrm{MHz}$. These values are obtained by translating our dimensionless operating point
and are comparable to those used in experimentally demonstrated superconducting
Duffing-resonator\text{---}In particular, ref.~\cite{chen2023NC} reports
resonator frequencies of $6.80\text{--}7.15~\mathrm{GHz}$, Kerr
nonlinearities of $U_{\mathrm{exp}}/2\pi=-295$ to $-58~\mathrm{kHz}$
in the convention $U_{\mathrm{exp}}a^{\dagger 2}a^2$, and an
energy-dissipation rate $\gamma=3.85~\mu\mathrm{s}^{-1}$, corresponding
to $\gamma/2\pi\simeq0.613~\mathrm{MHz}$. The associated dimensionless
temperature can be converted to laboratory units according to
$T_{\mathrm{phys}}=(\hbar\omega_c/k_{\mathrm B})(T/\omega_c)$.
For $\omega_c/2\pi=7.15~\mathrm{GHz}$, the interval
$T/\omega_c=0.1\text{--}0.3$ corresponds approximately to
$T_{\mathrm{phys}}=34\text{--}103~\mathrm{mK}$. This
temperature range is accessible in dilution-refrigerator circuit-QED
experiments \cite{blais2021RMP,scigliuzzo2020PRX,saira2016PRApp}.

\section{Conclusion}\label{Con}


In conclusion, we have investigated steady-state quantum thermometry in a driven--dissipative Kerr cavity coupled to a thermal reservoir. The system exhibits a finite-size precursor of a dissipative phase transition, identified by a pronounced minimum of the Liouvillian gap and a sharp crossover in the steady-state photon number. We have shown that this critical region, where the steady state is highly sensitive to temperature variations, gives rise to a strong enhancement of the quantum Fisher information for temperature estimation. Optimization over the coherent drive reveals that the enhanced response persists over an extended low-temperature, low-thermal-occupation regime.


The physical origin of the QFI enhancement has been clarified using an effective two-branch description of the steady state. Near the photon-number crossover, small changes in the bath temperature induce a strong redistribution of weight between the low- and high-photon-number branches. This temperature-sensitive branch redistribution provides the dominant population contribution to the QFI and is captured by the effective binary Fisher information
\[
F_{\rm bin}(F,T)=
\frac{[\partial_T p(F,T)]^2}{p(F,T)[1-p(F,T)]}.
\]
The agreement between the full steady-state QFI and the two-branch prediction shows that the enhancement is mainly governed by the temperature dependence of the effective branch occupation.

 Throughout the main numerical analysis, we have used $\omega_c=1$ as the frequency unit and choose $\kappa=0.08$, $U=0.03$, and $\Delta=0.6$. The ratios $U/\kappa=0.375$ and $\Delta/U=20$ place the cavity in a weakly nonlinear regime supporting a well-resolved many-photon crossover between low- and high-occupation responses. This parameter regime is convenient for elucidating how rapid nonequilibrium steady-state restructuring enhances thermometric sensitivity. These parameters, however, may be regarded as a benchmark rather than a representation of a specific device. Their non-uniqueness is illustrated by the comparison between the results for $\Delta=0.6$ and $\Delta=0.65$, which shows that varying the detuning shifts the photon-number crossover, minimum-gap region, and optimal-QFI line while preserving their qualitative association. Likewise, increasing the nonlinearity to $U/\kappa\sim 1$ or above changes the photon-number scale, crossover position, and QFI magnitude, but the enhanced thermometric response remains valid.




The close correspondence among the photon-number crossover,
Liouvillian-gap minimum, and QFI maximum confirms that the
mechanism identified in the main analysis---enhanced thermometric
sensitivity arising from rapid steady-state restructuring near a
finite-size precursor of a dissipative phase transition---persists in an experimentally realistic superconducting-circuit parameter regime, as illustrated in Fig.~\ref{fig:exp data}. 

These results show that finite-size dissipative-transition precursors in driven nonlinear cavities can enable tunable, experimentally accessible nonequilibrium quantum thermometry at low temperatures. Thus, the present work opens a route toward low-temperature quantum thermometry.

\textbf{Acknowledgments---} 
This work is supported by the National Quantum Mission (NQM), Department of Science and Technology, Government of India, under the Quantum Sensing and Metrology vertical.

\section{Appendix}

\appendix

\makeatletter
\@addtoreset{figure}{section}
\makeatother

\renewcommand{\thefigure}{\thesection\arabic{figure}}

\section{Effective two-branch description}
\label{app:two_branch}

\subsection{Representative temperature cuts of the two-branch diagnostics}
\begin{figure*}[t]
\includegraphics[width=1\textwidth]{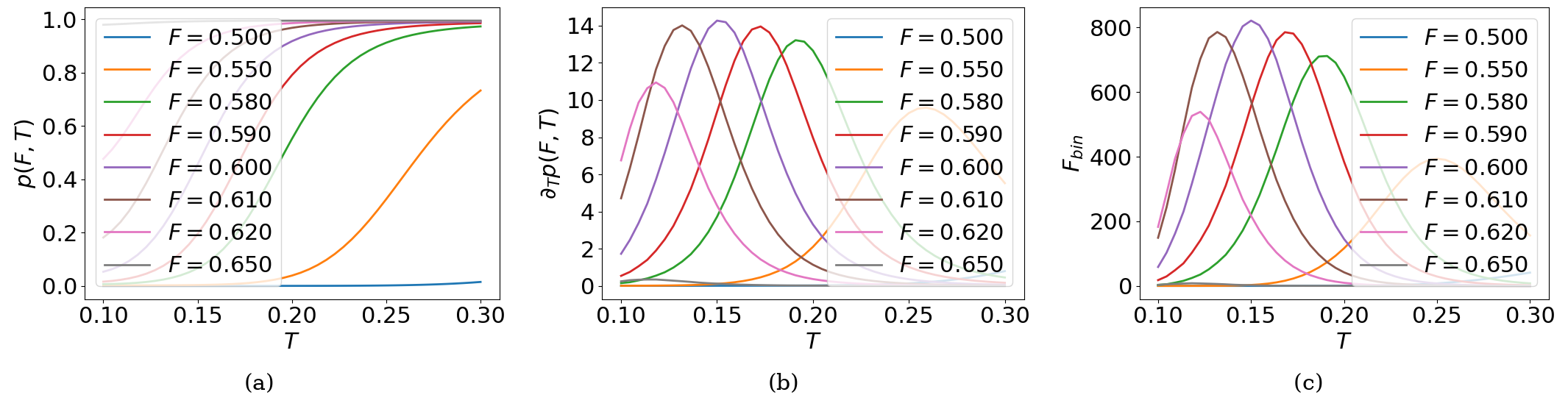}
   \caption{ Representative temperature cuts of the effective two-branch
    diagnostics for selected values of the drive amplitude \(F\).
    (a) Reconstructed occupation \(p(F,T)\) of the high-photon-number
    branch.
    (b) Corresponding temperature derivative
    \(\partial_T p(F,T)\).
    (c) Effective binary Fisher information
    \(F_{\rm bin}(F,T)\).
    The remaining parameters are the same as those used in Fig.~\ref{fig:Liouvillian Gap Vs Drive Amplitude}. } 
   \label{fig:branch-occupation-diagnostics-line-plots}
\end{figure*}

To complement the drive--temperature maps presented in Fig.~\ref{fig:branch_occupation_diagnostics}, we show in Fig.~\ref{fig:branch-occupation-diagnostics-line-plots} representative temperature cuts of the effective two-branch quantities for several fixed drive amplitudes. Fig.~\ref{fig:branch-occupation-diagnostics-line-plots}(a) shows the reconstructed high-photon-number branch occupation $p(F,T)$. For drive amplitudes intersecting the crossover region, $p(F,T)$ varies rapidly with temperature, reflecting a pronounced redistribution of the steady-state weight between the low- and high-photon-number branches. The corresponding derivative $\partial_Tp(F,T)$, shown in Fig.~\ref{fig:branch-occupation-diagnostics-line-plots}(b), develops peaks at the temperatures where this redistribution is most rapid. As a consequence, the binary Fisher information shown in Fig.~\ref{fig:branch-occupation-diagnostics-line-plots}(c) exhibits pronounced peaks in the same temperature regions. The shift of these peaks with drive amplitude illustrates that the temperature of maximum sensitivity can be tuned through the coherent drive.

\subsection{Two branch prediction without scaling factor}

\begin{figure}[h!]
\includegraphics[width=0.45\textwidth]{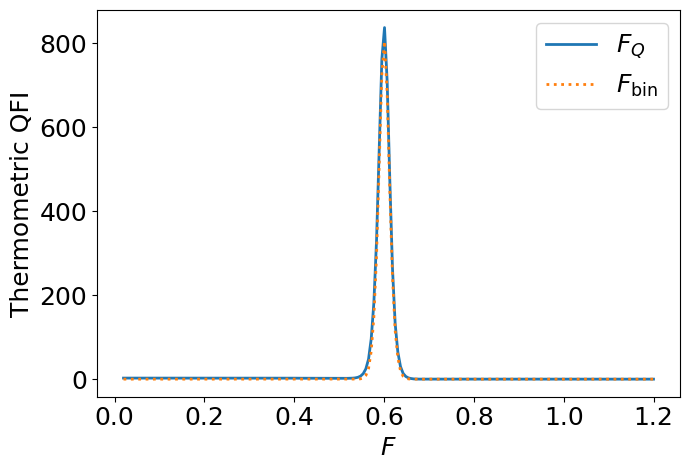}
   \caption{Variation of quantum fisher information $F_Q$ as a function of drive amplitude $F$. The other parameters are the same with Fig.~\ref{fig:Liouvillian Gap Vs Drive Amplitude}. } 
   \label{fig:Two State Mechanism Comparison}
\end{figure}

In Fig.~\ref{fig:Two State Mechanism Comparison}, we compare the full steady-state quantum Fisher information with the two-branch prediction Eq.~\ref{TBP} without applying any rescaling. While the overall magnitude of the two-branch prediction differs from that of the full QFI, the peak position and functional dependence are already well captured. In particular, the location of the QFI maximum and the narrow peak structure coincide closely in both curves.


The remaining discrepancy in amplitude arises from contributions beyond the idealized two-branch description, including residual temperature dependence of the effective branch states, coherence effects, branch deformation, and deviations from a perfect binary reduction. This comparison demonstrates that the two-branch model correctly captures the dominant mechanism underlying the thermometric enhancement, while the rescaling used in the main text, Fig.~\ref{fig:QFI Vs F with two state prediction}, emphasizes the agreement in the peak position and profile rather than the absolute magnitude.


\subsection{Convergence with Hilbert-Space Cutoff}

To ensure that the enhancement of the steady-state temperature QFI is not a numerical artifact of the finite Fock-space truncation,
we performed a convergence analysis with respect to the Hilbert-space cutoff \(N\). For each value of \(N\), the steady state
\(\rho_{\rm ss}(T)\) was computed using the same physical parameters, and the temperature QFI was evaluated from the mixed-state expression
\[
F_Q(T)=2\sum_{m,n}
\frac{|\langle m|\partial_T\rho_{\rm ss}|n\rangle|^2}
{p_m+p_n},
\] 
where \(\rho_{\rm ss}=\sum_m p_m |m\rangle\langle m|\), and \(\partial_T\rho_{\rm ss}\) was evaluated using a central finite difference in temperature.

\begin{figure*}[t]
\includegraphics[width=0.7\textwidth]{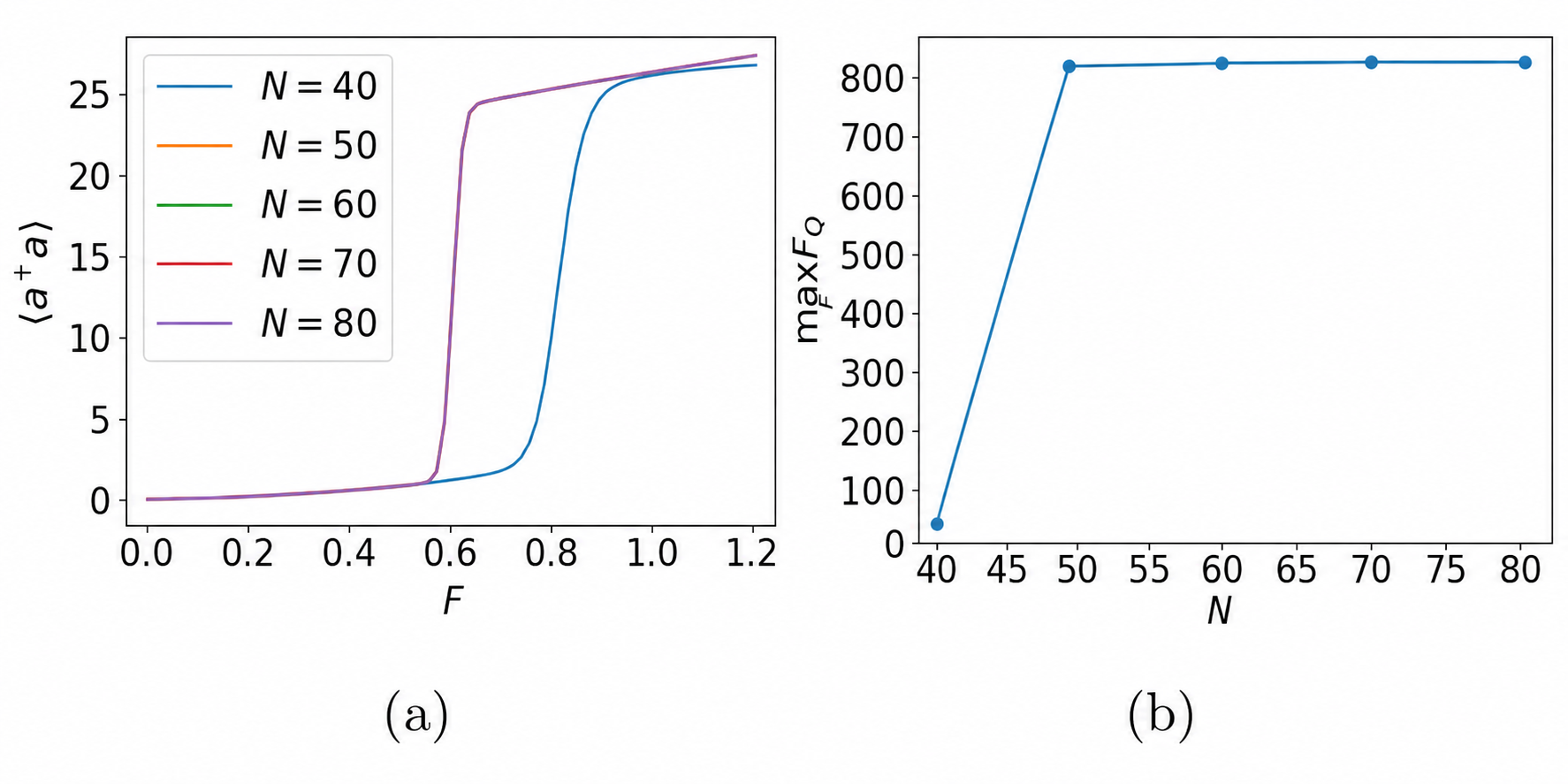}
   \caption{ (a) Steady-state photon number $\langle a^\dagger a\rangle_{\rm ss}$ as a function of the drive amplitude $F$ for $T=0.15$. (b) Peak value of the steady-state quantum Fisher information $F_Q^{\max}$ as a function of the Fock-space cutoff $N$. The other parameters are the same with Fig.~\ref{fig:Liouvillian Gap Vs Drive Amplitude}. } 
   \label{fig:Hilbert-Space Cutoff Convergence}
\end{figure*}

Fig.~\ref{fig:Hilbert-Space Cutoff Convergence}(a) shows the steady-state photon number as a function of the
drive amplitude, confirming that the chosen cutoff captures the relevant
low- and high-photon-number branches across the crossover region. In
Fig.~\ref{fig:Hilbert-Space Cutoff Convergence}(b), we plot the peak value of the QFI as a function of the cutoff
\(N\). The peak QFI changes appreciably for small cutoffs but saturates
beyond a sufficiently large value of \(N\). This saturation indicates
that the reported thermometric enhancement is converged with respect to
the Hilbert-space truncation. Therefore, the QFI peak near the finite-size
precursor of the dissipative phase transition is a genuine feature of the
driven--dissipative Kerr dynamics rather than a cutoff-induced artifact.

\section{Trace-Distance Diagnostic of Nonequilibrium Steady-State Character}\label{Trace dist}

To verify whether the enhanced thermometric response originates from ordinary thermalization, we compute the trace distance between the nonequilibrium steady state and the thermal state associated with the bath. The trace distance provides an operational measure of distinguishability between quantum states~ \cite{nielsen2010Book,fuchs1999book},
\[
D_{\rm th}
=
\frac{1}{2}
\left\|
\rho_{\rm ss}
-
\rho_{\rm th}^{\rm bath}
\right\|_1,
\]
where
\[
\rho_{\rm th}^{\rm bath}
=
\sum_{n=0}^{N-1}
\frac{\bar n^n}{(1+\bar n)^{n+1}}
|n\rangle\langle n|,
\qquad
\bar n =
\frac{1}{e^{\omega_c/T}-1}.
\]
In the weak-driving regime, the trace distance remains close to zero, indicating that the steady state is approximately thermal with respect to the bath. As the drive approaches the transition precursor, however, the trace distance increases sharply and becomes close to unity. This shows
that the steady state near the QFI peak is strongly distinguishable from the bath thermal state. Therefore, the enhancement of the steady-state QFI is not a trivial consequence of equilibration, but rather arises from the nonequilibrium restructuring of the steady state near the Liouvillian
gap minimum.


\begin{figure}[h!]
\includegraphics[width=0.45\textwidth]{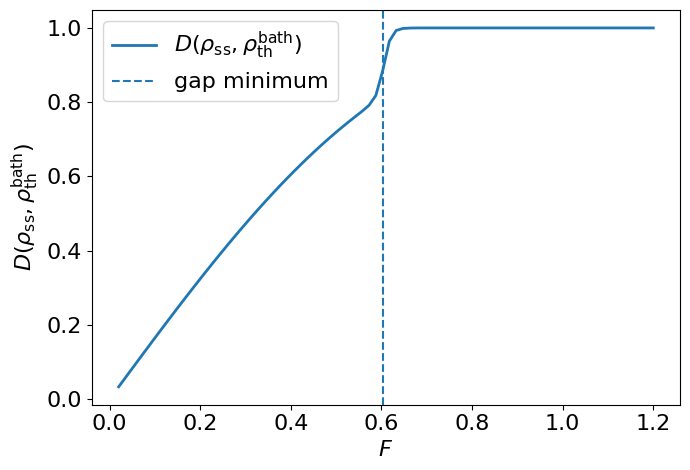}
\caption{
The trace distance \(D_{\rm th}=\frac{1}{2}\|\rho_{\rm ss}-\rho_{\rm th}^{\rm bath}\|_1\)
between the driven steady state and the bath thermal state is shown as a function of
the drive amplitude \(F\). The other parameters are the same with Fig.~\ref{fig:Liouvillian Gap Vs Drive Amplitude}. 
} 
\label{fig:Steady State}
\end{figure}

\bibliography{ref_new}

\end{document}